# Analysis of Various Transformer Structures for High Frequency Isolation Applications

Mohamad Saleh Sanjarinia, Sepehr Saadatmand, Pourya Shamsi, and Mehdi Ferdowsi
Department of Electrical and Computer Engineering
Missouri University of Science and Technology
Rolla, Missouri, USA
Corresponding Author: Mohamad Saleh Sanjarinia (mswvq@mst.edu)

*Abstract*— **High frequency transformers are an integral part of power electronics devices and their parasitic parameters influence the performance and efficiency of the overall system. In this paper, transformer leakage inductances and parasitic capacitances are analyzed using finite element method (FEM) for different structures and windings arrangements of high frequency transformers. Also, magnetic field, electric field, and voltage distribution within the transformer is simulated and analyzed. Six different high frequency transformers with toroidal, EE, and UU cores with different windings are investigated for a 400(V)/400(V), 8 kVA transformer operating at 10 kHz. Additionally, interleaved windings for EE core are simulated and results compared with previous outcomes. Analysis results will help categorize each structure, based on its balance between leakage inductances and series parasitic capacitance. This information can later be used for optimal selection of transformers as a function of their operating frequency and enable designers to compromise between various parameters in different applications, especially new fast switches such as SiC and GaN.**

*Index Terms*— **High Frequency Transformers, Leakage Inductance, Magnetic and Electric Field Distribution, Parasitic Capacitance, Winding Arrangements**

## I. INTRODUCTION

PARASITIC parameters in power electronics devices play an important role in device performance and efficiency. Many applications in power electronics use magnetic components. Because of advantages such as compactness, active/reactive power flow control, and protection, [1,2], HF galvanic isolation-based topologies have been widely incorporated in power electronics applications.

The leakage inductance in transformers results from imperfect coupling between the primary and the secondary of transformer windings which leads to voltage spikes, higher rectification losses, and efficiency reduction [3]. Additionally, parasitic capacitances lead to the injection of high frequency currents which increase the electromagnetic interferences (EMI). Moreover, parasitic capacitances may form electrostatic coupling to other parts of the circuit. Decreasing these two parameters not only improve switching waveforms, but also improve systems' efficiency.

Estimating the value of these parasitic parameters, can be very helpful during the design process. In some applications such as soft-switching converters, it is possible to use parasitic parameters of transformers as the additional inductors needed to form LC tanks. In [4], using leakage inductance of transformer as the inductance of tank circuit has reduced the total volume of the system by 15%.

As the frequency increases, the size of passive components, decreases. Hence, increasing the frequency can help the miniaturization of the converter. However, at higher frequencies, the importance of the parasitic, intensifies. For instance, the utilization of resonant soft-switching converters at higher frequencies rely on an accurate estimation of the parasitic in the final product. But unwanted leakage inductances and parasitic capacitances may interrupt resonant modes and produce unwanted resonant frequencies, which can distort voltage and current waveforms and decrease the efficiency. Also, this may lead to the interruption in control system [5].

In the design process of DC-DC converters, high frequency transformer design is an important step. Depending on the application and the design method, designers should decide to choose a suitable core shape and winding formation to deal with the parasitic parameters. Different transformer core shapes and winding arrangements, lead to different parasitic parameters. Therefore, it is necessary to understand the parasitic behavior of various transformer cores and ways of reducing and changing those parameters.

In this paper, six different transformer structures are investigated for high frequency and high power applications. The transformers are 1:1 and working at 400V for both primary and secondary at 10 kHz for power rating of 8 kVA. The windings are wound with AWG 8 (with diameter of 3.2639 mm) By using FEM magnetic field, electric field, and voltage distributions are computed for each structure then leakage inductance and parasitic capacitance for each of them are obtained. Then, by using interleaved windings and changing the voltage distribution along the cores, the parasitic parameters changed and the results compared with previous computations.

## II. TRANSFORMER DESIGN AND CORE SELECTION

Selecting magnetic material is a main step in designing high frequency transformers for power electronics applications. Cost, size, and performance are three major factors, which determine type of magnetic material in each design. These factors are influenced by relative permeability, temperature stability, core loss, cost, etc. Many magnetic materials can be used for the power electronics devices; and designers need to make a trade-off in choosing materials for each specific application. In TABLE I five magnetic materials are compared. It is noted that each feature is totally dependent to specific

TABLE I
Magnetic Materials key features

| | Molyperm alloy (MPP) | Sendust | Iron Powder | Ferrite MnZn | Ferrite NiZn |
|---|---|---|---|---|---|
| Temperature Stability | Very Good | Very Good | Very Good | Fair | Fair |
| Relative Permeability | 14-550 | 26-125 | 4-100 | 750-15000 | 15-1500 |
| Core Loss | Very Low | Low | Moderate | Very Low | Low |
| Relative Cost | High | Low | Very Low | Very Low | Very Low |

manufacturer. In this table ferrites are soft iron and the others are powder materials.

For the high frequency transformer in this paper, iron powder is chosen, mainly due to the lower frequency of application. Iron powder has a very low cost which make it suitable for mass production and shows very good temperature stability. Fig. 1 represent the B-H curve of the Iron powder used in this paper.

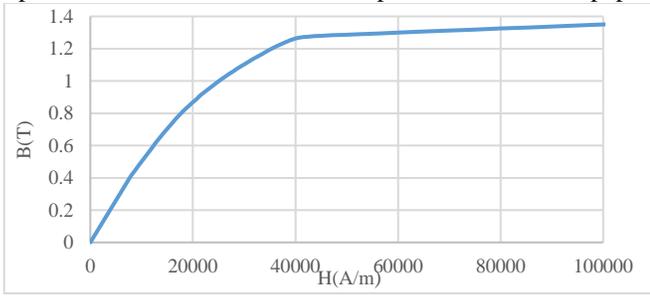

Fig. 1. Iron Powder-Mix-08 B-H curve used in the simulations for the cores.

By using (1), and following the design steps in [6], number of turn for both primary and secondary in all the transformers is set on 80 turns and the cross sectional area in all of them is 1600 mm$^2$. The turn ratio and cross sectional for all transformer structures set on the same values in order to make a better comparison, between the structures and transformers parasitic behavior.

$$P_{apparent} = K_f K_u B_m J f A_p \quad (1)$$

In (1), $P_{apparent}$, $K_f$, $K_u$, $B_m$, $J$, $f$, and $A_p$ are apparent power, waveform coefficient, window utilization factor, flux density, current density, and area product respectively.

To find core loss density the Steinmetz equation is widely used. However, [6] represents a more accurate formula for micro metals, which can be used for Iron powder-Mix-08. In (2), $P$ is the average of power loss density (watts/kg), $B$ is the amplitude of the flux density, $f$ is the system frequency, D is the duty cycle in square waveforms (D is 0.5 for sinusoidal waveforms) and $a$, $b$, $c$, $d$ are constants, which related to each material. For a specific material, at higher frequency, medium frequency and lower frequency values of these constants are different. These values for the iron powder-Mix-08 used here, are shown in TABLE IV.

$$P = \frac{fB^3(10^9)}{a+681(b)B^{0.7}+2.512(10^6)(c)B^{1.35}(c)} + 100(d)f^2B^2 \cdot \left(\frac{1}{D} + \frac{1}{1-D}\right) \cdot \frac{1}{4} \quad (2)$$

TABLE II
Constant values of the iron powder-Mix-08

| a | b | c | d |
|---|---|---|---|
| 0.01235 | 0.8202 | 1.4694 | 3.85×10$^{-7}$ |

### III. TRANSFORMERS STRUCTURES

Toroidal, UU, and EE cores are widely used, in the power electronics applications for the following reasons. The geometry of toroidal cores forces the magnetic field lines to form in closed circular loops and constrain the magnetic flux inside the core and reduce the EMI. Also, toroidal cores offer less weight and volume, and since they have a lower length of copper, the winding resistance decreases and due to large surface area of these core heat removal via convection and radiation is better in comparison. But winding copper wires on toroidal cores is more expensive than bobbin-based cores. Also, placing airgaps on the toroidal cores is quite difficult [7]. UU cores are also popular because of their simplicity, and high window utilization factor [7]. In some applications by using CC cores, which have rounded corners, high density of electric field can be removed. Lastly, EE cores are widely used in the power electronics applications. If airgaps are needed, EE cores allow forgapping of only the center leg which will reduce the EMI and fringing issues that arise when all legs are gapped [7]. Cooling of EE and UU cores are easier than toroidal cores due to their structure, which allows for a flow of coolant through the windings if winding spacers are used. Since, the windings in the EE cores are not completely covered by magnetic material, there is more EMI in these cores, which renders them not suitable for higher frequency applications. In such scenarios, pot cores can replace EE cores.

Fig. 2 shows the 3D transformer cores used for the simulations to obtain the parasitic parameters in this paper. Fig. 3 and TABLE III show the toroidal cores dimensions. Two toroidal cores types are investigated in this paper with different dimensions.

As it is shown in Fig. 2(a) and (b) both UU and EE cores are made with eight block sets. Fig. 4 and TABLE IV represent the block set and front and side views of the UU and EE cores. Moreover, in fig. 5 the winding arrangements used for the cores are depicted. The toroidal core with winding arrangements of same turns on the secondary of the toroidal core are wound in

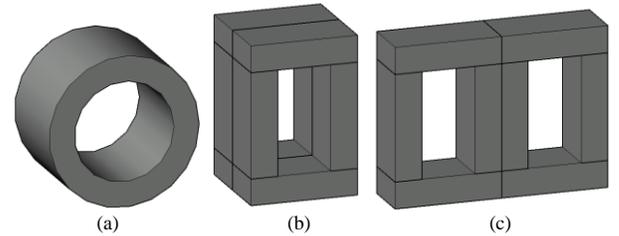

(a) (b) (c)
Fig. 2. The 3D transformers cores. a) Toroidal core. b) UU core. c) EE core

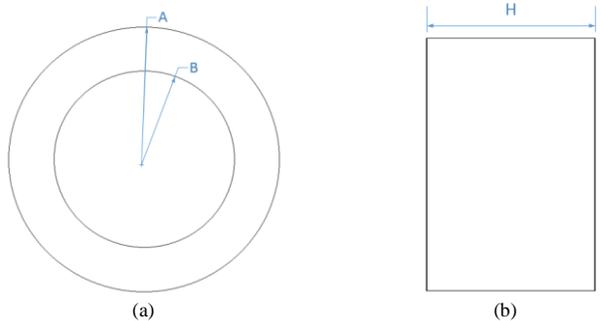

(a) (b)
Fig. 3. a) Front view of the toroidal core. b) Side view of the toroidal core (shows the depth of the core).

## TABLE III
### Dimensions of the toroidal cores

| | | |
|---|---|---|
| Case 1 | A | 60 (mm) |
| | B | 40 (mm) |
| | H | 80 (mm) |
| Case 2 | A | 70 (mm) |
| | B | 30 (mm) |
| | H | 40 (mm) |

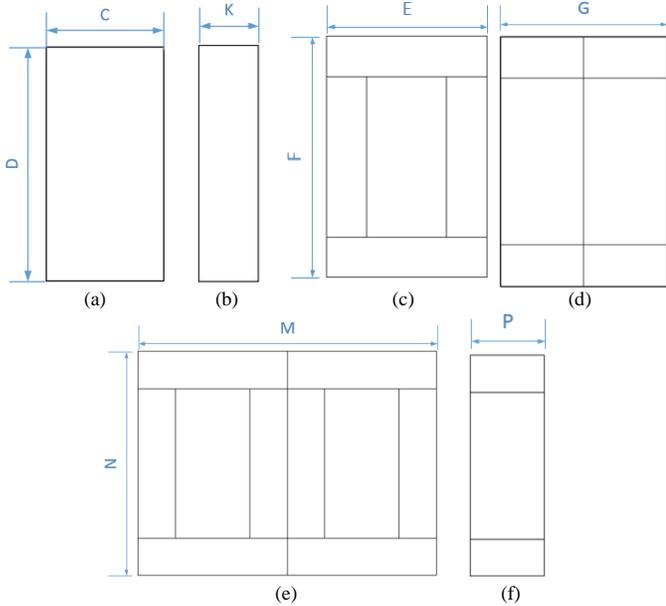

Fig. 4. Side and front views of the UU and the EE cores. a) Front view of the block set. b) Side view of the block set. c) Front view of the UU core. d) Side view of the UU core. e) Front view of the EE core. f) Side view of the EE core. Side views show the depth of the cores.

## TABLE IV
### Dimensions of the shapes of the Fig. 4.

| | |
|---|---|
| C | 40 (mm) |
| D | 80 (mm) |
| K | 20 (mm) |
| E | 80 (mm) |
| F | 120 (mm) |
| G | 80 (mm) |
| M | 160 (mm) |
| N | 120 (mm) |
| P | 40 (mm) |

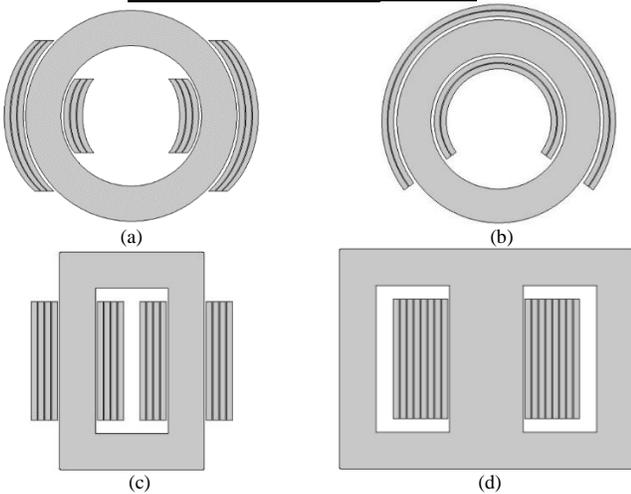

Fig. 5. Winding arrangements of the transformers. a) Toroidal core with 3 layers for each winding (non-overlaid windings). b) Toroidal core with 1 layer for each winding (overlaid windings). c) UU core winding arrangement with 4 layers for each winding. d) EE core winding arrangement with 4 layers for each winding.

three layers which is clear in Fig. 5(a). But for the toroidal core of Fig. 5(b) there is just one layer of 80 turns for the primary and the same for the secondary. This winding arrangements and layers are expected to completely influence the parasitic parameters. Furthermore, for the UU and the EE cores, 80 turns of the primary and the same turns of the secondary are wound in four layers for each of them.

## IV. PARASITIC PARAMETERS CALCULATIONS

In this section magnetic field, electric field, and voltage distributions for all the transformer cores, depicted in the section III, are displayed. Then, leakage inductance and parasitic capacitance for each structure calculated. Leakage inductance in transformers calculated by measuring energy and magneto motive force (MMF) stored in the space, while the secondary is short circuited and the parasitic capacitance calculated is between primary and secondary winding.

In high frequency, skin effect is the tendency of currents in conductors to concentrate on surface and by considering this phenomenon the leakage inductance values decrease. For accuracy purpose the skin effect is observed in the FEM simulations.

### A. Magnetic Field Distribution and Leakage Inductance Calculations

Fig. 6 shows magnetic flux density distributions, for all the cores. The flux density in this figure are obtained when the transformers work at the full load. As it is clear in this figure, the maximum flux density, in case 2 with 3 layers and case 2 with 1 layer are slightly greater than case 1 of both of 3 and 1 layers as well as flux density distribution linearity. In toroidal cores there is no saturation but in both UU and EE cores some points are in saturation. The winding arrangements of the layers for the EE core here, is the same as Fig. 5(d) with arrangement of P-P-P-P-S-S-S-S (P for primary and S for secondary) from left to right in the right window and a symmetry arrangement for the left window.

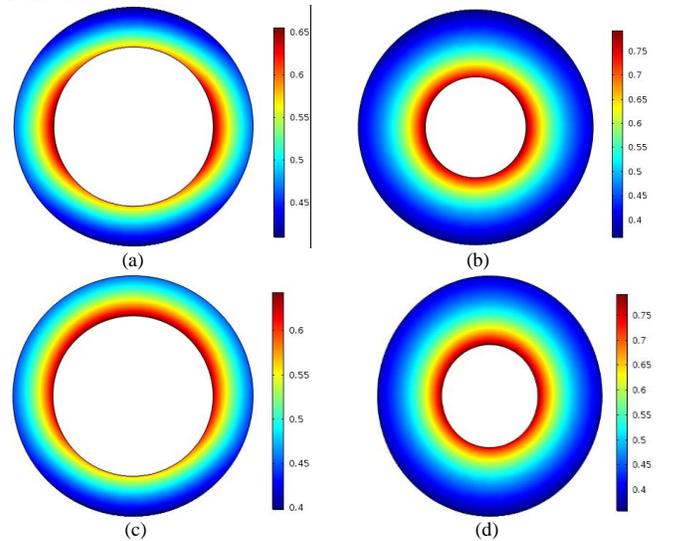

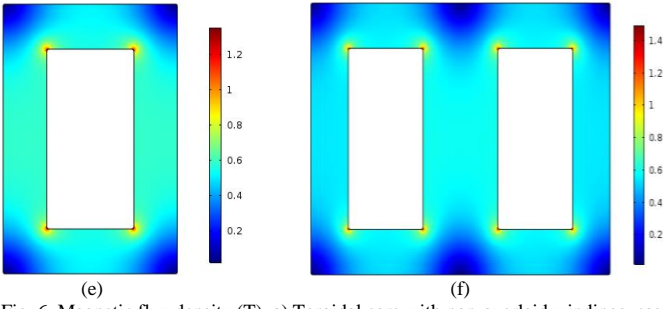
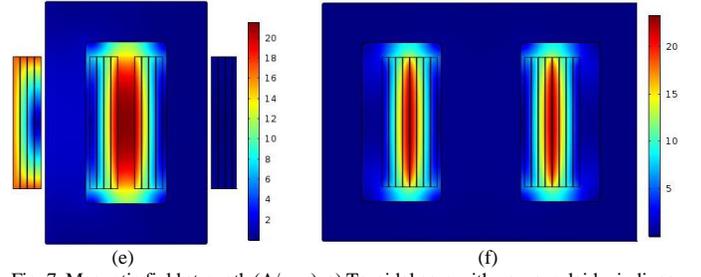

(e) (f)

Fig. 6. Magnetic flux density (T). a) Toroidal core with non-overlaid windings, case 1. b) Toroidal core with non-overlaid windings, case 2. c) Toroidal core with overlaid windings, case 1. d) Toroidal core with overlaid windings, case 2. e) UU core with 4 layers on each winding. f) EE core with 4 layers on each winding.

Fig. 7. Magnetic field strength (A/mm). a) Toroidal core with non-overlaid windings, case 1. b) Toroidal core with non-overlaid windings, case 2. c) Toroidal core with overlaid windings, case 1. d) Toroidal core with overlaid windings, case 2. e) UU core with 4 layers on each winding. f) EE core with 4 layers on each winding with the same winding arrangements of Fig. 6(f).

By short circuiting the secondary, MMF distribution at the space, especially the window area, is the key factor to find and analysis of leakage inductance. Hence, Leakage inductance is directly proportional to the energy stored in the space and with equations (4) and (5) magnetic energy in the space is related to the magnetic field strength (H).

$$MMF = \oint H . d\ell \quad (3)$$
$$E_{magnetic\ energy} = \frac{1}{2} \iiint H . B dv \quad (4)$$
$$B = \mu H \quad (5)$$
$$E_{total\ stored\ energy} = \frac{1}{2} . L_{leakage} . I_{primary}^2 \quad (6)$$

In the equations (3-6), $\ell$ is the length, $v$ is the volume of the space that energy is calculated on, $\mu$ is the permeability of the medium, $L_{leakage}$ is the leakage inductance from primary side, and $I_{primary}$ is the current of primary winding.

Fig. 7 represents the magnetic field strength (H) distribution in the transformer cores, winding arrangements, and window area while the secondary windings are short circuited. It is noted that besides the cores, the relative permeability of other parts such as insulations, copper of windings, and air are considered 1. But the topology and structures of the transformers influence the leakage inductance values. In these cores, the windings in all layers are wound homogeneously and the insulation between bare areas of the winding conductors is fixed on 0.15mm for all the topologies. Also, insulation and bobbin thickness between the windings and the core in the toroidal, UU and the EE cores is 2mm.

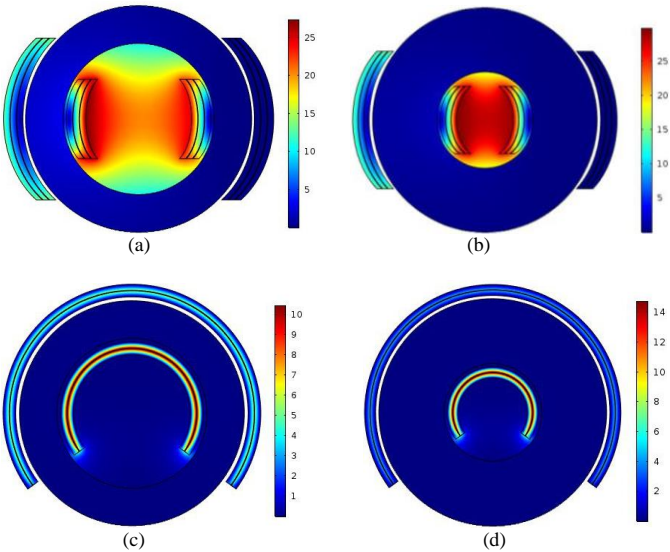

TABLE V
Magnetizing inductance, leakage inductances, and windings AC resistance of the transformers

| Transformer Type | Magnetizing Inductance($\mu$H) | Leakage Inductance($\mu$H) | AC Resistance(m$\Omega$) |
|---|---|---|---|
| a | 2061.5 | 736.25 | 34.055 |
| b | 1944.0 | 324.89 | 18.343 |
| c | 1774.5 | 15.21 | 50.817 |
| d | 1820.0 | 9.23 | 25.496 |
| e | 2160.8 | 450.42 | 36.306 |
| f | 2005.1 | 79.39 | 25.038 |

The leakage inductance and the magnetizing inductance for all the cores are achieved by FEM and displayed in TABLE V. Results obtained by considering magnetic field strength not only in the represented areas of Fig. 7, but also in the whole space. Also, in this table AC resistance of the windings from primary side are shown. The magnetizing inductance of the transformer types purposely designed to be almost near to each other, to make a better comparison between the parasitic parameters and choose the suitable type for each application. a,b,c,d,e,f types are the same as Fig. 6 and Fig. 7.

It is clear in TABLE V, leakage inductance can vary in a wide range by changing core and arrangements. By comparing leakage inductances of types a, b, c, and d and considering magnetic field strength of Fig. 7, it is apparent that the window area volume and the winding arrangements, which lead to better magnetic coupling are remarkably important in the results.

*B. Insulation Designs, Voltage Distributions, and Capacitance Calculations*

Parasitic capacitance between primary and secondary windings can be found by calculating total energy stored in the electric field. Then, by having total amount of stored energy between the primary and the secondary, capacitance can be calculated. Equations (7-9) explain these relations.

$$E_{electric\ energy} = \frac{1}{2} \iiint D . E dv \quad (7)$$
$$D = \varepsilon E \quad (8)$$
$$E_{total\ stored\ energy} = \frac{1}{2} C . V^2 \quad (9)$$

In equations (7-9), $v$ is the volume of the space that energy is calculated on, $\varepsilon$ is the permittivity of medium, $V$ is the related voltage, and $C$ is the capacitance.

The capacitance between the primary and the secondary is influenced by structure of transformers, winding arrangements,

and permittivity of used materials. The insulation dimensions between layers, windings, cores, and bobbin thickness, for this study are the same as mentioned before. Insulation material between bare conductors is silicon vanish with relative permittivity of 3.1 and the bobbin between windings and core are made from plastic with relative permittivity of 2.2.

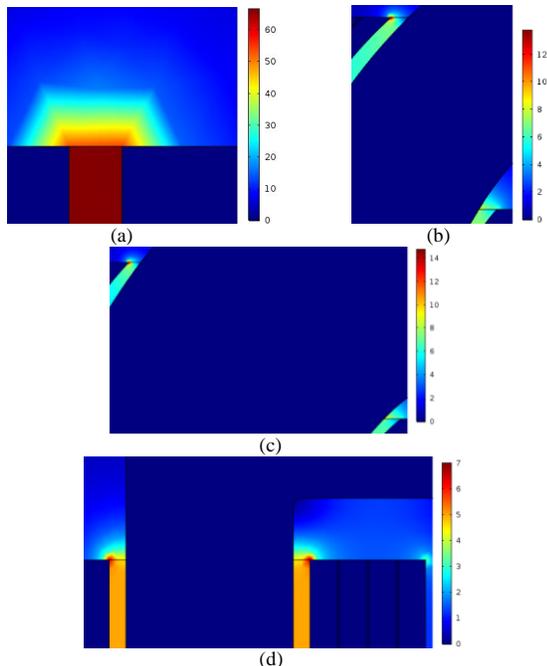

Fig. 8. Electric field distribution (kV/mm) in the transformers at highest insulation test voltage of 10kV. a) EE core, case 1, and case 2 of the toroidal cores with overlaid windings (these cores follow a similar maximum electric field pattern). Maximum electric field of 67(kV/mm) occurs between primary and secondary layers of these cores. b) Case 1 of the toroidal core with non-overlaid windings. Maximum electric field of 13.8(kV/mm) occurs between primary layers and the core. c) Case 2 of the toroidal core with non-overlaid windings. Maximum electric field of 14.7(kV/mm) occurs between primary layers and the core. d) UU core. Maximum electric field of 7(kV/mm) occurs between primary layers and the core.

From insulation design point of view, in addition to dielectric constant (relative permittivity) and distances, dielectric strength of the insulators are also important. The dielectric strength for silicon varnish and plastic of bobbins are considered 120(kV/mm) and 25(kV/mm) respectively, also air dielectric strength is considered 3(kV/mm). It is important to attention that the dielectric constants and strength for these materials can vary, because of different manufacturers and this may change the capacitance. According to the with IEEE Std. C57.12.01 dry-type transformers standard, which used in [8], the insulation designs of this study transformers, should be based on 10kV. To examine if the insulation materials and dimensions satisfy the standard or not; electric field distribution for the transformers computed with FEM. In this distribution, the primary is connected to the high insulation test voltage of 10kV and both secondary and cores are grounded. As it is shown in Fig. 8 maximum electric field in insulations is much lower than the dielectric strength of the used materials.

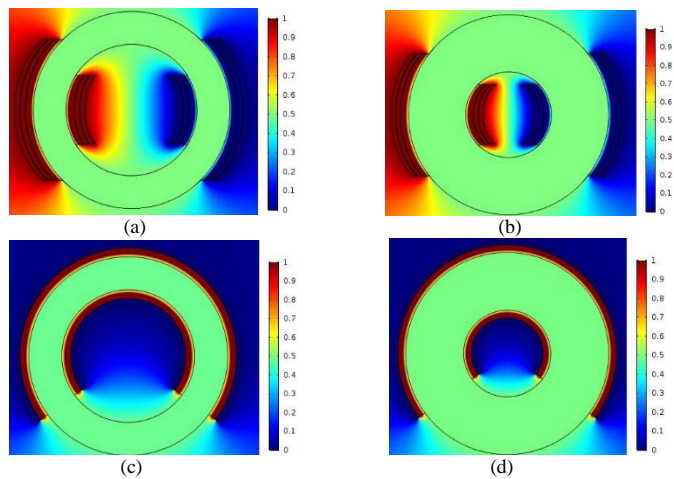

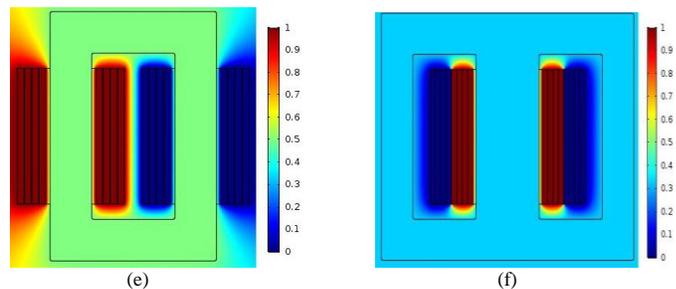

Fig. 9. Voltage distribuiton in the transforemrs by applying 1V to a winding and geounding the other one. a) Toroidal core with non-overlaid windings, case 1. b) Toroidal core with non-overlaid windings, case 2. c) Toroidal core with overlaid windings, case 1. d) Toroidal core with overlaid windings, case 2. e) UU core with 4 layers on each winding. f) EE core with 4 layers on each winding.

TABLE VI
Parasitic capacitacne(pF) between primary and secondary windings by considering two core types

| Transformer Type | Iron Powder-Mix-08 | Ferrite-NiZn |
|---|---|---|
| a | 76.258 | 1.869 |
| b | 30.603 | 1.346 |
| c | 6444.137 | 6444.278 |
| d | 3222.014 | 3222.269 |
| e | 60.577 | 7.3117 |
| f | 960.103 | 958.872 |

In order to find capacitance between windings by following procedure used in [9], 1V applied to the primary windings and 0V to the other ones. Fig. 9 shows voltage distribution in the transformer structures. Also, TABLE VI displays capacitance between primary and secondary windings the transformers. Since the conductivity of the core influence capacitance values, in this table the values by considering ferrite-NiZn as the core (with a very low conductivity) are also mentioned. a,b,c,d,e,f types are the same as Fig. 9.

As it is clear in TABLE VI parasitic capacitacne can change hugely for different structures. Additionally, by substituting a conductive core (Iron Powder) with a nonconductive core (Ferrite-NiZn with $10^5 (\Omega.cm)$ resistivity) the capacitacne can dramatically change for some cases.

### C. Interleaved Windings

By using interleaved windings, instead of the winding arrangements explained in the previous part, leakage inductance significantly decrease. For the EE core here, using interleaved windings reduce leakage energy stored in the window area and decrease MMF. Thus, without changing magnetizing inductance, leakage inductance decreases. As well, by using interleaved windings, parasitic capacitance changes too. Fig. 10 shows magnetic field strength distribution for three interleaved winding arrangements (while the secondary is short circuited) for the EE core. TABLE VII display new leakage inductance and parasitic capacitance for these windings. a,b,c, types are the same as Fig. 10.

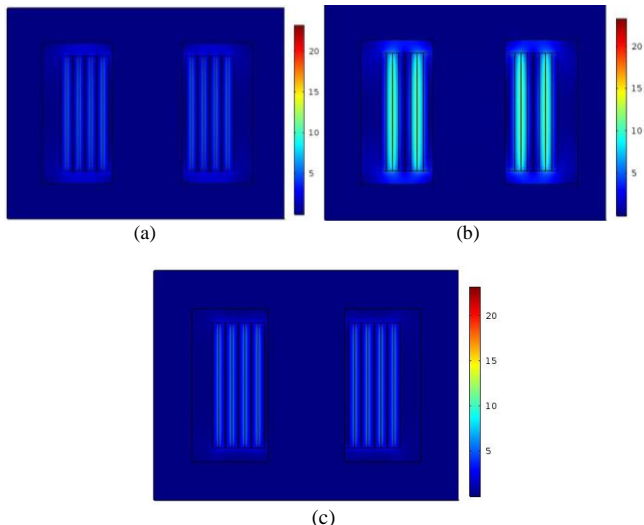

(a)  (b)

(c)

Fig. 10. Magnetic field strength (A/mm) for EE core by interleaved windings. These Color diagrams are scaled to be compared with Fig. 7(f). a) Interleaved winding arrangements of P-S-P-S-P-S-P-S from left to right in the right window and a symmetry arrangement for the left window. b) Interleaved winding arrangements of P-P-S-S-P-P-SS from left to right in the right window and a symmetry arrangement for the left window. c) Interleaved winding arrangements of P-S-S-P-P-S-S-P from left to right in the right window and a symmetry arrangement for the left window.

TABLE VII
Leakage inductance and parasitic capacitance of the EE core with interleaved windings

| Interleaved EE Transformer Type | Leakage Inductance($\mu$H) | Parasitic Capacitance(pF) |
|---|---|---|
| a | 5.12 | 6704.1 |
| b | 19.90 | 2875.0 |
| c | 5.51 | 3832.3 |

The maximum of magnetic field strength in both Fig. 10(a) and (c) reaches to about 6 (A/mm) and in Fig. 10(b) it reaches to almost 11.8 (A/mm), while in the Fig. 7(f) with P-P-P-P-S-S-S-S arrangement the maximum is 23.3 (A/mm). However, magnetic field, leakage stored energy in the window area and consequently leakage inductance for this transformer decreases significantly by interleaved windings, the parasitic capacitance increases significantly. Additionally, in TABLE VII type a and b have almost the same leakage inductance but a completely different capacitance.

## V. CONCLUSION

By comparing the results of the structures and the winding arrangements, it can be concluded that by changing design parameters a wide range of leakage inductance and parasitic capacitance can be obtained. Depend on the application and utilization, all of the studied structures can be used for the high frequency transformers. To change the leakage inductance changing the magnetic field strength distribution is necessary because it leads to changing magnetic stored energy. By this method or using different structures, the leakage inductance for high frequency transformers can change and by reduction of that, the efficiency can increase. But this may lead to a different parasitic capacitance between primary and secondary windings which in some applications with fast switches (SiC and GaN) with high $\frac{dV}{dt}$ result to injecting high frequency current that may create many troubles for the system. In some structures like EE cores using interleaved windings, may be a good solution to reduce leakage inductance, although this leads to a higher parasitic capacitance. Finally, to choose a structure and a specific design strategy, each application and system should be considered and analyzed separately, and between leakage inductance and parasitic capacitance a trade-off is vital. Also, manufacturing factors in each application are important and may impose some design compromises.

## ACKNOWLEDGEMENT

This material is based upon work supported by the U.S. Department of Energy, "Enabling Extreme Fast Charging with Energy Storage", DE-EE0008449.

## REFERENCES

[1] H. Qin and J. Kimball, "Ac-ac dual active bridge converter for solid state transformer," *2009 IEEE Energy Conversion Congress and Exposition*, 2009.
[2] B. Zhao, Q. Song, W. Liu, and Y. Sun, "Overview of Dual-Active-Bridge Isolated Bidirectional DC–DC Converter for High-Frequency-Link Power-Conversion System," *IEEE Transactions on Power Electronics*, vol. 29, no. 8, pp. 4091–4106, 2014.
[3] Z. Ouyang, J. Zhang, and W. G. Hurley, "Calculation of Leakage Inductance for High-Frequency Transformers," *IEEE Transactions on Power Electronics*, vol. 30, no. 10, pp. 5769–5775, 2015.
[4] J.-M. Choi, B.-J. Byen, Y.-J. Lee, D.-H. Han, H.-S. Kho, and G.-H. Choe, "Design of Leakage Inductance in Resonant DC-DC Converter for Electric Vehicle Charger," *IEEE Transactions on Magnetics*, vol. 48, no. 11, pp. 4417–4420, 2012.
[5] S. Yazdani and M. Ferdowsi, "Robust Backstepping Control of Synchronverters under Unbalanced Grid Condition," *2019 IEEE Power and Energy Conference at Illinois (PECI)*, 2019.
[6] C. W. T. McLyman, *Transformer and inductor design handbook*. CRC Press, 2017.
[7] M. K. Kazimierczuk, *High-frequency magnetic components*. Chichester: John Wiley & Sons, 2014.
[8] S. Zhao, Q. Li, F. C. Lee, and B. Li, "High-Frequency Transformer Design for Modular Power Conversion From Medium-Voltage AC to 400 VDC," *IEEE Transactions on Power Electronics*, vol. 33, no. 9, pp. 7545–7557, 2018.
[9] M. B. Shadmand and R. S. Balog, "A finite-element analysis approach to determine the parasitic capacitances of high-frequency multiwinding transformers for photovoltaic inverters," *2013 IEEE Power and Energy Conference at Illinois (PECI)*, 2013.